# Electron correlation energy in confined two-electron systems

C.L. Wilson[a], H.E. Montgomery, Jr.[a,*], K. D. Sen[b], D. C. Thompson[c]

[a]Chemistry Program, Centre College, 600 West Walnut Street, Danville, KY 40422 USA

[b]School of Chemistry, University of Hyderabad, Hyderabad 500 046, India

[c]Chemistry Systems and High Performance Computing, Boehringer Ingelheim Pharamaceuticals Inc., 900 Ridgebury Road, Ridgefield, CT 06877 USA

**Abstract**

Radial, angular and total correlation energies are calculated for four two-electron systems with atomic numbers Z = 0 – 3 confined within an impenetrable sphere of radius *R*. We report accurate results for the non-relativistic, restricted Hartree-Fock and radial limit energies over a range of confinement radii from 0.05 – 10 $a_0$. At small *R,* the correlation energies approach limiting values that are independent of Z while at intermediate *R*, systems with Z ≥ 1 exhibit a characteristic maximum in the correlation energy resulting from an increase in the angular correlation energy which is offset by a decrease in the radial correlation energy.

* Corresponding author.
*E-mail address:* ed.montgomery@centre.edu (H.E. Montgomery).

## 1. Introduction

The concept of electron correlation energy ( $E_{corr}$ ) was first proposed by Wigner [1] in his classic study of free electrons in a metal. It was subsequently discussed by Gell-Mann and Brueckner [2] in the context of an electron gas at high density and later defined by Löwdin [3] as the difference

$$E_{corr} = E - E_{HF}\,, \tag{1}$$

where $E$ is the exact, non-relativistic energy, $E_{HF}$ is the limiting unrestricted Hartree-Fock energy and $E_{corr}$ is seen to be negative.

Correlation energy is among the most important and difficult problems in quantum chemistry, having been characterized by Tew et al. [4] as "The many-body problem at the heart of chemistry." It is important because calculations of chemical accuracy require accurate evaluation of the correlation energy and because correlation energy calculations are a testing ground for the exchange potentials of density functional theory. It is difficult because accurate calculation of the correlation energy as a difference requires calculation of the non-relativistic energy and the Hartree-Fock energy to very high accuracy.

Understanding the factors that determine the correlation energy is complicated by the dearth of parameters that can be conveniently varied to measure their effect. For two-electron systems, the obvious parameter is $Z$, the nuclear charge. The effects of varying $Z$ have been studied in great detail by Koga et al [5]. Ezra and Berry [6] constrained the electrons to move on the surface of a sphere and investigated Coulombic, Gaussian and $\delta$-function repulsive interactions. They then extended this work to two particles on concentric spheres [7]. Recently Loos and Gill have varied $D$, the dimensionality of two-electron systems and evaluated the resulting changes in the correlation energy. They considered *D*-helium, *D*-spherium [8] where the electrons move in a constant potential on the surface of a hypersphere of radius $1/Z$, *D*-hookium [9,10] where the electrons move in the harmonic potential $V(r) = r^2/2$ and *D*-ballium [11,12] where the electrons are confined by an impenetrable barrier in a *D*-dimensional ball of radius *R*. Such a set of model systems are useful in providing further insight to the electron correlation problem.

The present study focuses on the effects of changing the radius of confinement on the correlation energy of ballium, the $H^-$ ion, He and the $Li^+$ ion confined at the center of an impenetrable sphere. The correlation between the electrons can be varied by changing the size of the system.

This work follows from the pioneering work of Gimarc [13], who investigated the correlation energy of the confined, two-electron atom using a three-term, explicitly correlated wavefunction to calculate the non-relativistic energy and a double-zeta wavefunction [14] to calculate the Hartree-Fock energy. Recent work using explicitly correlated wavefunctions to calculate accurate ground state energies [15-18] and accurate unrestricted Hartree-Fock calculations provided an opportunity to perform a detailed study confined two-electron systems. To the best of our knowledge, these are the first high-accuracy correlation energy calculations performed on confined atomic systems since the work of Gimarc.

We note here that the correlation energy within the Kohn-Sham (KS) model [19] of the density functional theory (DFT) [20] is defined in a different manner. The total KS energy-density *functional* is prescribed as the sum of the kinetic energy $T_s[\rho]$ for a noninteracting-electron system, the electron–external-potential interaction energy, the electron-electron classical Hartree Coulomb repulsion $E_h[\rho]$, and $E_{xc}[\rho]$, the exchange-correlation energy. Thus, all non-classical electron interactions, i.e. Pauli exchange, electron correlation, and $T_c[\rho]$ -the difference between the kinetic energy of the interacting- and non-interacting-electron systems, are represented by the unknown functional $E_{xc}$ [ρ], the functional derivative of which defines the exchange-correlation potential. Accurate quantitative evaluation of these various energy contributions using the electron density, ρ(r), derived from the wave functional calculations have been reported earlier [21] for the He-isoelectronic series , among other light atoms, in the unconfined state.

The accurate investigation of the properties of the various model systems listed above, with their tunable parameters for investigating correlation, provide an excellent opportunity for systematic exploration, and improvement, of the myriad density functionals currently available to the electronic structure [22,23].

The non-relativistic energies were calculated using explicitly correlated expansions in Hylleraas coordinates [24] while the Hartree-Fock calculations were accomplished through a *B*-spline approach, which was independently verified through use of a method employing a spherical Bessel function representation [25].

As pointed out by Taylor and Parr. [26], the correlation energy can be partitioned into the radial correlation energy $\left(E_{rad,corr}\right)$ and the angular correlation energy $\left(E_{ang,corr}\right)$. Radial correlation accounts for the tendency of the two electrons to be at different distances from the nucleus while angular correlation accounts for the tendency of the two electrons to occupy positions on opposite sides of the nucleus. Following the method of Koga [27], the radial limit energy was calculated by modifying the Hylleraas expansion to exclude angular correlation from the wavefunction. The radial correlation energy was then obtained as

$$E_{rad,corr} = E_{radial} - E_{HF} . \qquad (2)$$

The angular correlation energy was found by subtracting $E_{radial}$ from $E$ to obtain

$$E_{ang,corr} = E - E_{radial} \,. \tag{3}$$

A principal aim of the present work was to investigate the variation of $E_{rad,corr}$ and $E_{ang,corr}$ caused by changes in *Z* and in the radius of confinement, *R*.

## 2. Computational details

Calculation of the non-relativistic energies followed the method of ten Seldam and de Groot [30] through the development of the secular determinant. Atomic units were used throughout. The explicitly correlated wavefunction for two electrons confined in an impenetrable sphere of radius *R* was an expansion of the form

$$\psi = \left[R - \frac{1}{2}(s-t)\right]\left[R - \frac{1}{2}(s+t)\right] e^{-\alpha s} \sum_{k=1}^{N} c_k s^{l_k} t^{m_k} u^{n_k} \,, \tag{4}$$

where *s*, *t* and *u* are the Hylleraas coordinates [19] defined by

$$s = r_1 + r_2; \quad t = -r_1 + r_2; \quad u = r_{12} \,. \tag{5}$$

The factors $\left[R - \frac{1}{2}(s \pm t)\right]$ are cutoff functions which insure that the wavefunction goes to zero at *R*. The $c_k s$ are variational parameters determined by solving the secular determinant while the non-linear parameter $\alpha$ was determined by hand minimization. Since the exponential $e^{-\alpha s}$ incorporates the electron-nuclear attraction, $\alpha$ for ballium was set equal to zero and the wavefunction had the same form as equation (21) of [7]. We note in passing that for $R \leq 1\ a_0$, setting $\alpha = 0$ introduces error in sixth decimal place, even for $Z = 2$.

All terms with $l_k + m_k + n_k \leq 7$ were included in the wavefunction, subject to the requirement that $m_k$ = even. This gave a 70-term wavefunction. For the radial energy, we used a similar wavefunction but required $n_k = 0$ and included all terms with $l_k + m_k \le 15$. This resulted in a 72-term wavefunction.

The required integrals were evaluated as the sum of the three sets of integrals

$$\int ds\,dt\,du = \int_0^R ds \int_0^s du \int_0^u dt + \int_R^{2R} ds \int_0^{2R-s} du \int_0^u dt + \int_R^{2R} ds \int_{2R}^{s} du \int_0^{2R-s} dt \tag{6}$$

The calculations were coded in Maple and performed on a 2.66 GHz dual-core processor. Typical computational time was 70 minutes.

After the completion of these calculations, we became aware of the work of Pan et al. [29], who have developed a formulation of the Hylleraas integrals that changes the order of integration and thereby reduces equation (6) to two integrals. We have subsequently investigated the effect on the time of integration and have found that implementation of the method of Pan gives a 30% reduction in the time required to evaluate the Hylleraas integrals without affecting the numerical accuracy.

The restricted Hartree-Fock energies were calculated through a method employing a zeroth order spherical Bessel function representation of the Hartree-Fock orbitals. This method and its implementation has been described in detail in [25]. In a basis of zeroth order spherical Bessel functions, convergence with respect to principal quantum number ($\mu$) is rapid, requiring an $\mu_{max}$ of 7 for $10^{-9}$ $E_h$ accuracy for the Z=0 problem [7,25]. As was observed in [25], for Z $\neq$ 0 convergence of this basis is far slower with respect to $\mu$, and is a much subtler function of $R$. For $H^-$, He, and $Li^+$ several representative $R$ were investigated and converged with respect to $\mu$. We have also employed a $B$-spline approach in order to evaluate the performance of the zero order spherical Bessel representation. $B$-spline calculations were carried out using a modified HF code [16] with the choice of 100-term $B$-spline set of order K = 9. A preset finite radius of confinement $R$ on an exponential type knot sequence [17] with the initial interval as $10^{-4}$ was employed. The zero order spherical basis set Hartree-Fock calculations have been found to be in quantitative agreement with the $B$-spline basis results.

### 3. Results and discussion

$E$, $E_{radial}$ and $E_{HF}$ were calculated over the range of $R$ from 0.05 $a_0$ to 10 $a_0$. Selected energies and the corresponding correlation energies of equations (1-3) are shown in Table 1. Also shown is $\%E_{corr}$, defined as

$$\%E_{corr} = \left|\frac{E_{corr}}{E}\right| \times 100 . \tag{7}$$

The correlation energies are summarized in Figure 1 which shows $E_{rad,corr}$ and $E_{corr}$ as a function of $R$ with $E_{ang,corr}$ given by the distance between the curves. As $R$ approaches 0, $E_{rad,corr}$ and $E_{corr}$ for the $H^-$ ion, He and the $Li^+$ ion approach the energies for ballium. The limiting values were found to be $E_{rad,corr}$ = 0.0032 $E_h$, $E_{corr}$ = 0.0552 $E_h$ and $E_{ang,corr}$ = 0.0520 $E_h$.

For large $R$, approaching the free systems, $E_{corr}$ for $H^-$, He and $Li^+$ is relatively constant even though $E$ for $Li^+$ is 13 times greater than $E$ for $H^-$ with a resulting decrease in $\%E_{corr}$. This constancy of $E_{corr}$ results from a small decrease in $E_{rad,corr}$ which is offset by a similar increase in $E_{ang,corr}$. As pointed out by Gimarc [13], the electrons do most of their correlating of motion inside what we normally consider the dimensions of the atom or ion. A case could be made for defining the dimension of a system as that region outside of which the correlation energy becomes constant.

For ballium, $E_{rad,corr}$ is relatively constant, increasing from -0.003133 $E_h$ at $R$ = 0.05 $a_0$ to -0.0012755 $E_h$ at $R$ = 10 $a_0$. $E_{ang,corr}$ is ~15 times larger than $E_{rad,corr}$ and increases from -0.051816 $E_h$ at $R$ = 0.05 $a_0$ to -0.026821 $E_h$ at $R$ = 10 $a_0$. Thus $E_{corr}$ for ballium depends largely on angular correlation even at large $R$.

For the $H^-$ ion, $E_{corr}$ goes through three distinct regions as shown in Figure 2. For increasing $R$ at tight confinement, $E_{corr}$ is dominated by the increase in $E_{ang,corr}$, similar to what was seen for ballium. However, the presence of an attractive nucleus results in a concurrent decrease in $E_{rad,corr}$ which offsets the increase in $E_{ang,corr}$ resulting in a maximum in $E_{corr}$ at $R$ = 5.51 $a_0$. The decrease in $E_{rad,corr}$ can be thought of as resulting from the increase in confinement volume which increases the probability of the two electrons being found at different distances from the nucleus and thus stabilizes the system. $E_{rad,corr}$ and $E_{ang,corr}$ cross at $R$ = 7.21 $a_0$ with $E_{rad,corr}$ remaining lower than $E_{ang,corr}$ as $R$ increases. For $R$ > ~15 $a_0$, confinement has little effect and the correlation energies approach their free-system values.

Helium and $Li^+$ show similar dependence on $R$ except the maximum in $E_{corr}$ is shifted inward to $R$ = 2.73 $a_0$ for helium and to $R$ = 1.84 $a_0$ for $Li^+$. Also, the crossover between $E_{rad,corr}$ and $E_{ang,corr}$ was found to be unique to $H^-$, with $E_{ang,corr}$ remaining lower than $E_{rad,corr}$ at all values of $R$ for $Z \geq 2$.

**4. Conclusions**

In this work, we have reported accurate values for non-relativistic, radial limit and Hartree-Fock energies for confined two-electron systems with $Z$ = 0, 1, 2 and 3 and have used them to find the radial and angular contributions to the correlation energy over a range of values of nuclear charge and confinement. We find that the correlation energies are relatively constant except at very tight confinement, even though the system energies vary significantly.

Gimarc [13] conjectured that the partitioning of $E_{ang,corr}$ and $E_{rad,corr}$ with changing *R* would differ from the free-system values. The correctness of his conjecture can be easily seen in Figures 1 and 2. The behavior of the system is can be characterized in terms of the following observations.

1. $E_{ang,corr}$ is the primary contributor to $E_{corr}$ for ballium, $H^-$, He and $Li^+$ at tight confinement. This is consistent with our picture of tight confinement reducing the opportunities for radial correlation and most of the correlation energy resulting from inclusion of angular terms in the wavefunction. As *R* increases, $E_{ang,corr}$ for $H^-$, He and $Li^+$ increases monotonically with decreasing slope approaching the free-system value from below. As Z increases, the value of *R* at which free-system behavior is obtained decreases, consistent with decreased size of the atom/ion. Similarly, the free-system value of $E_{ang,corr}$ decreases as *Z* increases. We expect those trends to continue for Z > 3.

2. For $H^-$, He and $Li^+$, $E_{rad,corr}$ decreases with increasing *R*, approaching the free-system value from above. This is consistent with our picture of increased opportunity for the electrons to be at different distances from the nucleus as the volume of the system increases, The slope of the $E_{ang,corr}$ versus *R* curve is negative at small *R*, becomes increasingly negative as *R* increases, then passes through a minimum before increasing to zero for large *R*. These minima in the slope occur at $R = 5.32\ a_0$, $1.49\ a_0$ and $1.18\ a_0$ for $H^-$, He and $Li^+$ respectively. As *Z* increases, we expect the minima to shift to smaller *R*.

3. When $E_{ang,corr}$ and $E_{rad,corr}$ are added to give $E_{corr}$, the result is a maximum in $E_{corr}$ that occurs at a value of *R* that is slightly larger than the minima listed above. Based on the behavior of $E_{ang,corr}$ and $E_{rad,corr}$ discussed above, we expect the maximum in $E_{corr}$ to shift to smaller *R* and lower energy with increasing *Z*.

$E_{ang,corr}$ and $E_{rad,corr}$ are found as small differences between the non-relativistic, radial limit and Hartree-Fock energies. They are determined by the electron-nuclear attraction, the electron-electron repulsion and the requirement that the wavefunction go to zero at *R*. The detailed interactions behind the shapes of the curves are both interesting and complicated. Better understanding of the details of these interactions offers interesting possibilities for future work.

It is hoped that this careful investigation of correlation for a series of two-electron systems will be a boon to the electronic structure community, both through a provision of benchmark quality data, and through additional physical understanding of the phenomena of correlation dual functions of confining and nuclear potentials.

**Acknowledgements**

D.C.T. would like to thank Drs. Pitt and Alavi from the Cambridge University Centre for Computational Chemistry ($CUC^3$) for the use of computer resources. K.D.S. is grateful to Dr. S.L. Saito for a copy of the *B*-spline Atomic HF code and to Dr. C. Froese-Fisher for constant encouragement with the implementation of the *B*-spline basis. H.E.M acknowledges financial support by the Centre College Faculty Development Committee.

**References**

[1] E. Wigner, Phys. Rev. 46 (1934) 1002.

[2] M. Gell-Mann, K. A. Brueckner, Phys. Rev. 106 (1957) 364.

[3] P.-O. Löwdin, Adv. Chem. Phys. 2 (1959) 207.

[4] D.P. Tew, W. Klopper, T. Helgaker, J. Comp. Chem. 28 (2007), 1307.

[5] T. Koga, M. Omura, H. Teruya, A.J. Thakkar, J. Phys. B: At. Mol. Opt. Phys**.** 28 (1995) 3113.

[6] G.S. Ezra, R.S. Berry, Phys. Rev. A 25 (1982) 1513.

[7] G.S. Ezra, R.S. Berry, Phys. Rev. A 28 (1983) 1989.

[8] P.-F. Loos, P. M. W. Gill, Phys. Rev. Lett. 103 (2009) 123008.

[9] P.-F. Loos, P. M. W. Gill, J. Chem. Phys. 131 (2009) 241101.

[10] N. R. Kestner, O. Sinanoglu, Phys. Rev. 128 (1962) 2687.

[11] D. C. Thompson, A. Alavi, Phys. Rev. B 66 (2002) 235118.

[12] P.-F. Loos, P. M. W. Gill, J. Chem. Phys. 132 (2010). 234111.

[13] B.M. Gimarc Chem. Phys. 47 (1967) 5110.

[14] L.C. Green, M.M. Mulder, M.N. Lewis, and J.W. Woll Phys. Rev. 93 (1954) 757.

[15] N. Aquino, A. Flores-Riveros, J. F. Rivas-Silva, Phys. Lett. A 307 (2003) 326.

[16] N. Aquino, J. Garza, A. Flores-Riveros, J.F. Rivas-Silva, K. D. Sen, J. Chem. Phys. 124 (2006) 054311.

[17] A. Flores-Riveros, N. Aquino H. E. Montgomery, Jr., Phys. Lett. A 374, (2010) 1246.

[18] H.E. Montgomery Jr., N. Aquino, A. Flores-Riveros, Phys. Lett. A 374, (2010) 2044.

[19] P. Hohenberg, W. Kohn, Phys. Rev. 136 (1964) B864.

[20] W. Kohn, L. Sham, Phys. Rev. 140 (1965) A1133.

[21] C.J. Umrigar, X. Gonze, Phys. Rev. A 50 (1994) 3827; Chien-Jung Huang, C. J. Umrigar, Phys. Rev. A 56 (1997) 296.

[22] S. Kais, D.R. Herschbach, N.C. Handy, C.W. Murray, G.J. Lamming, J. Chem. Phys. 99 (1993) 417.

[23] J. Jung, P. Garcia-Gonzalez, J. E. Alvarellos, R.W. Godby, Phys. Rev. A 69 (2004) 052501.

[24] E. A. Hylleraas, Z. Phys. 54 (1929) 347.

[25] D. C. Thompson, A. Alavi, J. Chem. Phys. 122 (2005) 124107.

[26] G.R. Taylor, R.G. Parr, Proc. Natl. Acad. Sci. USA 38 (1952) 154.

[27] T. Koga, Z. fur Physik 37 (1996) 301.

[28] C. A. ten Seldam, S. R. de Groot, Physica 18 (1952) 891.

[29] X-Y Pan, V Sahni, L. Massa, arXiv:physics/0310128v3.

[30] S. L. Saito, J. Chem. Phys. 130 (2009) 074306; S. L. Saito, Atomic Data and Nuclear Data Tables 95 (2009) 836–870.

[31] T.L. Gilbert, P.J. Bertoncini, J. Chem. Phys. 61 (1974) 3026.

**Table 1.** Selected energies for confined $2e^-$, $H^-$, He and $Li^+$. $R$ is in $a_0$. Energies are in $E_h$

| | $R$ ($a_0$) | $E_{HF}$ | $E_{RL}$ [1] | $E$ [2] | $E_{rad,corr}$ | $E_{ang,corr}$ | $E_{corr}$ | $\%E_{corr}$ |
|---|---|---|---|---|---|---|---|---|
| $2e^-$ | 0.05 | 3983.548808 | 3983.545675 | 3983.493858 | -0.003133 | -0.051816 | -0.054950 | 0.0014 |
| | 1.0 | 11.641749 | 11.638807 | 11.590839 | -0.002942 | -0.047969 | -0.050910 | 0.4392 |
| | 5.0 | 0.739762 | 0.737467 | 0.701614 | -0.002295 | -0.035853 | -0.038148 | 5.4372 |
| | 10.0 | 0.266624 | 0.264870 | 0.238049 | -0.001755 | -0.026821 | -0.028575 | 12.0040 |
| $H^-$ | 0.05 | 3885.925658 | 3885.922469 | 3885.870899 | -0.003188 | -0.051570 | -0.054759 | 0.0014 |
| | 1.0 | 6.637526 | 6.633326 | 6.589644 | -0.004200 | -0.043682 | -0.047882 | 0.7266 |
| | 5.0 | -0.425815 | -0.438594 | -0.461974 | -0.012779 | -0.023380 | -0.036159 | 7.8270 |
| | 10.0 | -0.486150 | -0.509209 | -0.524688 | -0.023059 | -0.015478 | -0.038538 | 7.3449 |
| | ∞ | -0.487930 | -0.514489 | -0.527748 | -0.026560 | -0.013258 | -0.039818 | 7.5449 |
| He | 0.05 | 3787.859261 | 3787.856017 | 3787.804693 | -0.003245 | -0.051324 | -0.054569 | 0.0014 |
| | 1.0 | 1.061203 | 1.055135 | 1.015755 | -0.006067 | -0.039380 | -0.045448 | 4.4743 |
| | 5.0 | -2.861390 | -2.878668 | -2.903409 | -0.017278 | -0.024741 | -0.042019 | 1.4472 |
| | 10.0 | -2.861680 | -2.879025 | -2.903724 | -0.017345 | -0.024699 | -0.042044 | 1.4479 |
| | ∞ | -2.861680 | -2.879025 | -2.903724 | -0.017345 | -0.024699 | -0.042044 | 1.4479 |
| $Li^+$ | 0.05 | 3689.341971 | 3689.338669 | 3689.287592 | -0.003302 | -0.051077 | -0.054379 | 0.0015 |
| | 1.0 | -5.318324 | -5.326867 | -5.362399 | -0.008544 | -0.035531 | -0.044075 | 0.8219 |
| | 5.0 | -7.236415 | -7.252486 | -7.279912 | -0.016071 | -0.027426 | -0.043497 | 0.5975 |
| | 10.0 | -7.236415 | -7.252487 | -7.279913 | -0.016072 | -0.027426 | -0.043498 | 0.5975 |
| | ∞ | -7.236415 | -7.252487 | -7.279913 | -0.016072 | -0.027426 | -0.043498 | 0.5975 |

[1] 72 term Hylleraas expansion in $s$ and $t$

[2] 70 term Hylleraas expansion in $s$, $t$ and $u$

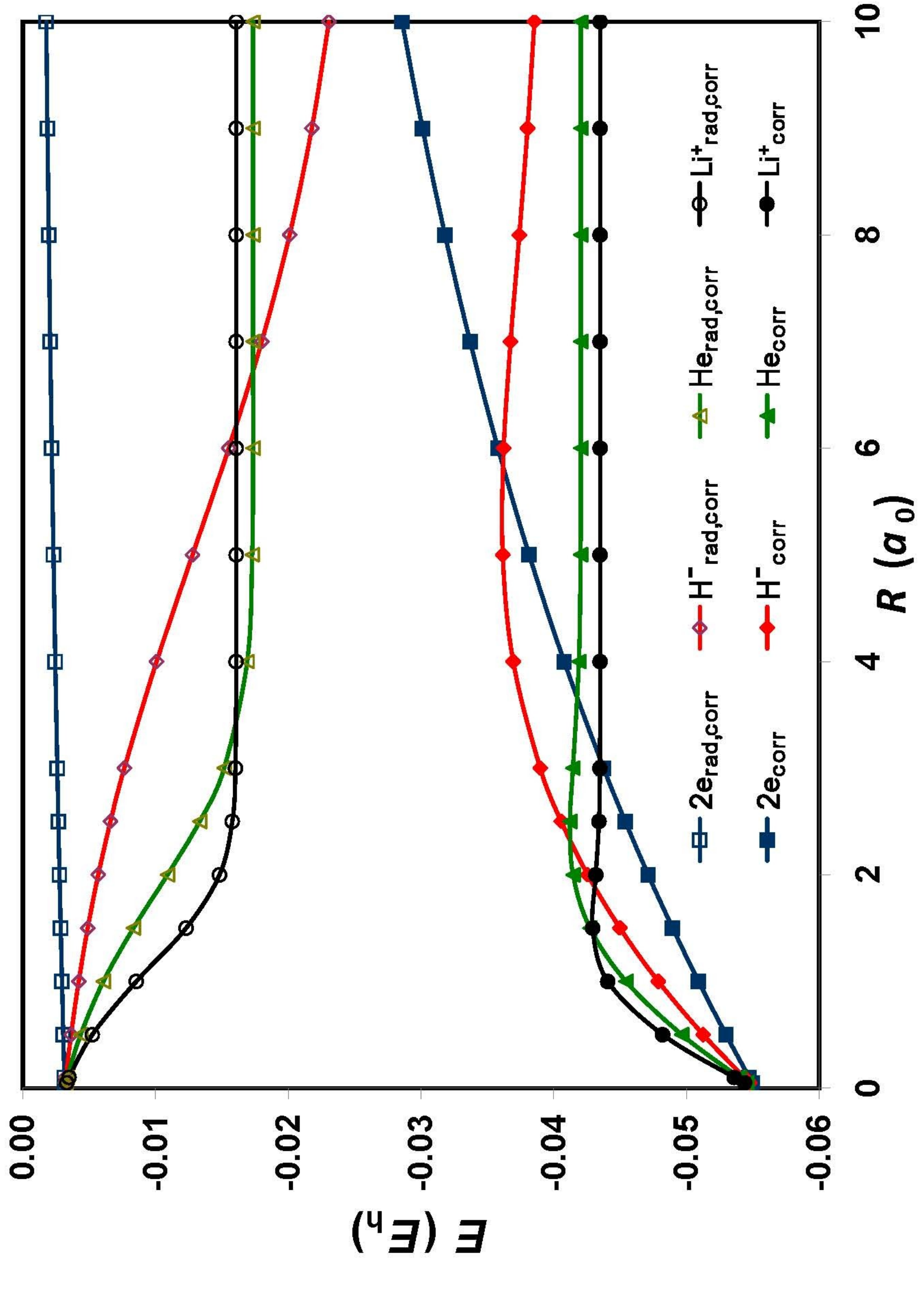

**FIG 1.** $E_{rad,corr}$ and $E_{corr}$ for 2e$^{-}$, H$^{-}$, He and Li$^{+}$

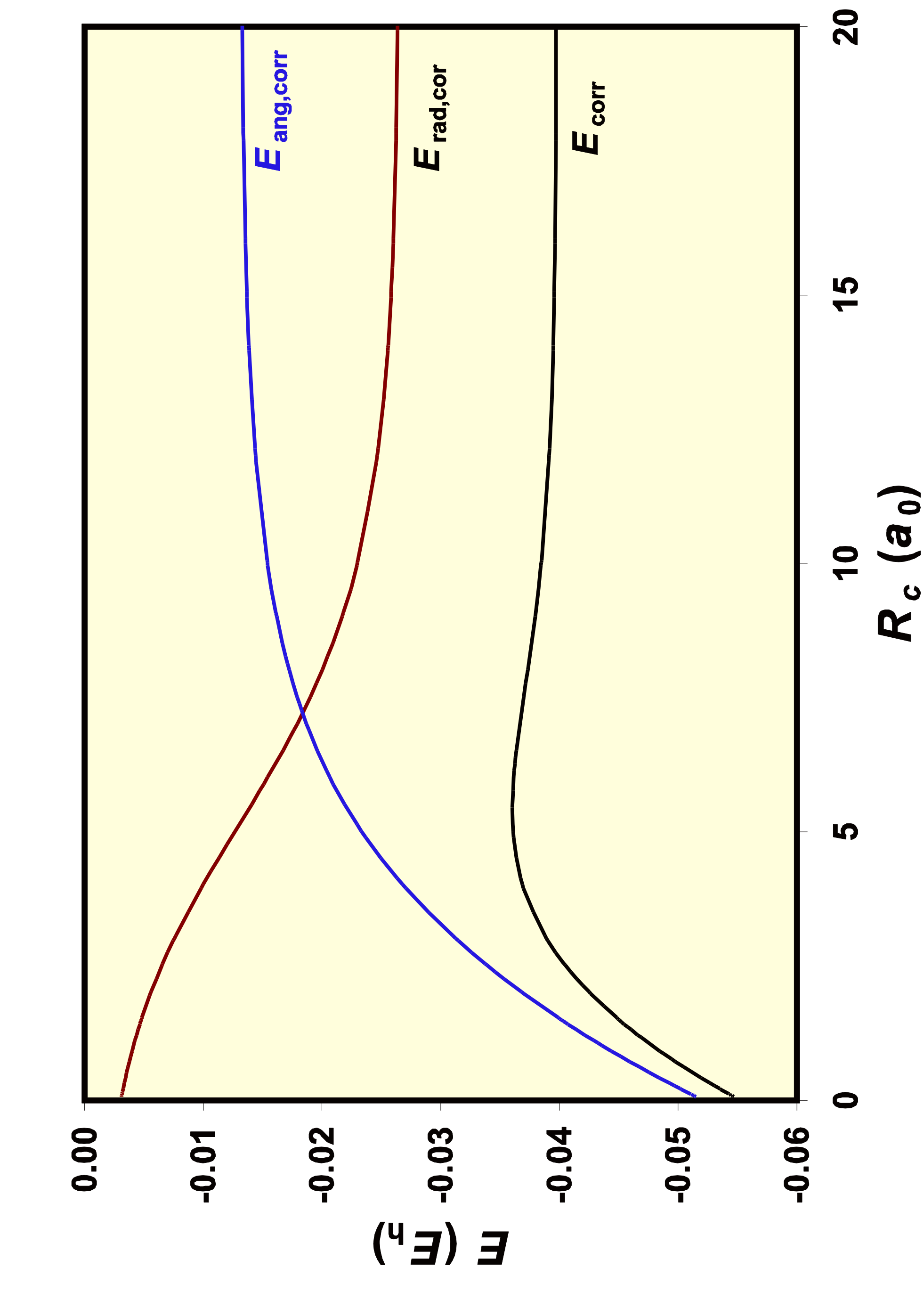

**FIG 2.** Correlation energies for $H^-$